\documentclass[letterpaper,titlepage,12pt]{article}
\usepackage{setspace} % Always before hyperref -- Enable correct footnote link
\usepackage[colorlinks=false,urlcolor=blue,citecolor=magenta]{hyperref}
\usepackage{xcolor}
\usepackage{amssymb,amsmath,amsfonts}
\usepackage{epsfig}
\usepackage{epsfig}
\usepackage{graphicx}
\usepackage{epstopdf}
\usepackage[round]{natbib}
\usepackage{caption}
\usepackage{subcaption}
\usepackage{amscd}
\usepackage{amsthm}
\usepackage{csquotes}
\usepackage{latexsym}
\usepackage{float}
\usepackage{mathtools}
\usepackage[stable]{footmisc}
\usepackage{amsbsy}
\usepackage{bbm}
\usepackage[english]{babel}
\usepackage{psfrag}
\usepackage{tabularx}
\usepackage{csquotes}
\usepackage{colortbl}
\usepackage{braket}

\allowdisplaybreaks

\newcommand{\be}{\begin{eqnarray}}
\newcommand{\ee}{\end{eqnarray}}
\newcommand{\beq}{\begin{eqnarray}}
\newcommand{\eeq}{\end{eqnarray}}

\def\clock{{\count0=\time
 \divide\count0 60
 \ifnum\count0<10 0\fi\the\count0
 \multiply\count0 -60 \advance\count0 \time
 :\ifnum\count0<10 0\fi \the\count0
 }}
\newcommand{\timestamp}{{\small\vbox{\hbox{\tt\jobname.tex}
\hbox{\the\day/\the\month/\the\year, \clock}}}}

\begin{document}

%\today

\vskip 1.4 cm
\centerline{\LARGE \bf Lack of Value Definiteness in Quantum Gravity}
\vskip .25cm
\vskip 1.5cm
\centerline{\large {{\bf Enrico Cinti$^{1,2}$, Cristian Mariani$^{3}$, Marco Sanchioni$^1$}}}
\vskip .8cm
\begin{center}
\sl $^1$ DISPeA, University of Urbino,\\
\sl Via Timoteo Viti 10, 61029 Urbino PU, Italy.\\
\end{center}
\begin{center}
\sl $^2$ Department of Philosophy, University of Geneva,\\
\sl 5, rue de Candolle, CH-1211 Gen\`eve 4, Switzerland.\\
\end{center}
\begin{center}
\sl $^3$ Institut Néel, CNRS (Grenoble), \\
25 Avenue des Martyrs, 38000 Grenoble, France.\\
\end{center}
\vskip 0.6cm
\centerline{\small\tt marco.sanchioni2@gmail.com, cinti.enrico@gmail.com, cristian.mariani@neel.cnrs.fr}

\vskip .8cm \centerline{\bf Abstract} \vskip 0.2cm \noindent In this paper we consider the observables describing fundamental spatiotemporal properties and relations in the context of Quantum Gravity (QG). As we will show, in both Loop Quantum Gravity and in String Theory, these observables are non-commuting ones. By analogy with some recent arguments put forward in the context of non-relativistic quantum mechanics (QM), we shall propose to interpret these physical quantities as \textit{ontologically indeterminate}---i.e., indeterminate in a non-epistemic, mind-independent way. This result has two important consequences for current debates in philosophy of physics. First, it shows that \textit{ontological indeterminacy} may extend well beyond the non-relativistic case, thereby also suggesting a conceptual continuity between QM and QG. Second, when applied to QG, the notion of indeterminacy may provide a unified philosophical framework for understanding very distinct approaches that are usually considered incompatible.

\noindent 


%%%%%%%%%%%%%%%%%%%%%%%%%%%%%%%%%%%%%%%%%%%%%%%%%%%%%%%%%%%%%%%%%%%%%%%%%%%%%%%%%%%%%%%%%%%%%%%%%%%%%%%%%%%%%%%%%%%%%%%%%%%%%%%%%%%%%%%%%%%%%%%%%%%%%%%%%%%%%%%%%%%%%%%%%%%%%%%%%%%%%%%%%%%%%%%%%%%%%%%%%%%%%%%%%%%%%%%%%%%%%%%%%%%%%%%%%%%%%%%%%%%%%%%%%%%%%%%%%%%%%%%%%%%%%%%%%%%%%%%%%%%%%%%%%%%%%%%%%%%%%%%%%%%%%%%%%%%%%%%

\newpage

\tableofcontents

%%%%%%%%%%%%%%%%%%%%%%%%%%%%%%%%%%%%%%%%%%%%%%%%%%%%%%%%%%%%%%%%%%%%%%%%%%%%%%%%%%%%%%%%%%%%%%%%%%%%%%%%%%%%%%%%%%%%%%%%%%%%%%%%%%%%%%%%%%%%%%%%%%%%%%%%%%%%%%%%%%%%%%%%%%%%%%%%%%%%%%%%%%%%%%%%%%%%%%%%%%%%%%%%%%%%%%%%%%%%%%%%%%%%%%%%%%%%%%%%%%%%%%%%%%%%%%%%%%%%%%%%%%%%%%%%%%%%%%%%%%%%%%%%%%%%%%%%%%%%%%%%%%%%%%%%%%%%%%%
\section{Introduction}\label{sec1}
As it is well known, in any quantum theory certain pairs of physical quantities, namely the \textit{non-commuting} ones, cannot simultaneously be assigned a definite value. This feature is at the core of many philosophical discussions regarding the ontology of the theory, its conceptual implications, and more generally its mathematical and physical foundations. In many ways, it could be argued that the \textit{lack of value definiteness} (henceforth, LVD) is the mark that distinguishes quantum theories from classical ones. In recent years, a novel view has emerged which attempts to interpret LVD in a more systematic way and by recognising how pervasive this feature is. The idea, which actually traces back to the fathers of quantum theory,\footnote{For instance, a section of \cite{schrodinger1935gegenwartige} is entitled \enquote{Are the variables really fuzzy?}.} is to take LVD at face value as indicating that the natural world is ontologically indeterminate in some respect, i.e. it is indeterminate independently from our knowledge of it, or from our representations. Ontological Indeterminacy (OI), a.k.a. metaphysical indeterminacy, has been proposed within the context of the Orthodox interpretation of quantum mechanics \citep{David2021}, and also, more recently, as an ontology for spontaneous collapse models \citep{mariani2020non}, for the Many-Worlds interpretation \citep{calosi2021quantumreview}, and for Relational Quantum Mechanics \citep{calosi2020quantum}. Common to all these proposals is the idea that we can individuate certain physical quantities that are crucial for describing the ontology of the given theory (such as \textit{position}, \textit{spin}, or \textit{mass}), and then show that they are ontologically indeterminate. 

Although the current discussion only focuses on examples from non-relativistic quantum mechanics, the mathematical and conceptual features on which the existing arguments rely should extend to virtually every theory that has the right to be called \textit{quantum}. In particular, a crucial role in this debate is played by the  \textit{non-commutativity} of the algebra of the operators \citep{calosi2019quantum,wolff2015spin}. Consequently, in this paper we shall investigate the status of LVD within two of the major approaches to Quantum Gravity (QG), namely Loop Quantum Gravity (LQG) and String Theory (ST).  

When it comes to LQG, the non-commutativity which interests us comes from attempting to reconcile the existence of a minimal area in the theory with the need for Lorentz invariance, which demands that all lengths vary continuously under Lorentz boosts. This reconciliation relies on Lorentz boosted operators not commuting with their unboosted counterparts. With regards to ST, non-commutativity is a consequence of the quantization of the scalar fields living on the string's worldsheet. We can then start by canonically quantizing the scalar fields of the string, which can be treated as a type of position variable. Together with their associated conjugate momenta, one finds that these scalar fields form a non-commutative algebra of operators, satisfying non-commutativity relations essentially analogous to those of the well known position/momentum case in quantum mechanics. The crucial point for this paper is that in both cases the consistency of the theory relies crucially on the claim that certain operators cannot represent determinate quantities at the same time. These properties are in LQG the geometric properties encoded by the area operator, while in ST they are centre of mass position, centre of mass momentum, and higher harmonic modes of the string. In LQG we have an analogy with angular momentum, which makes particularly explicit the connection with analyses of ontological indeterminacy already developed in the existing literature. In ST too, the fact that we are basically dealing with non-commutativity of position and momentum operators is analogous with the standard non-relativistic case.

The main difference between our case and the previously discussed ones—as we will show more extensively in this paper—concerns the way in which we could attribute physical meaning to these quantities, and the kind of properties that are involved. In the case of quantum mechanics (QM), the common way of assigning definite values to observables is \textit{via} the Eigenstate-Eigenvalue Link, which roughly establishes a one-to-one mapping between definite properties and eigenstates. In the context of QG, however, the status of the EEL is yet to be fully understood. Even more importantly, we shall notice that the existing arguments for \textit{ontological indeterminacy} based on the EEL show that certain \textit{monadic} properties are indeterminately instantiated. In QG, instead, it seems that the resulting indeterminacy affects \textit{relational} properties or structures.

\bigskip

\noindent \textit{Roadmap}. In \S\ref{sec2} we introduce the standard argument for the existence of \textit{ontological indeterminacy} in the non-relativistic QM case, and we then briefly introduce the main existing approaches to this issues. We then present the cases for the existence of quantum indeterminacy in Loop Quantum Gravity and String Theory, respectively in \S\ref{sec3} and \S\ref{sec4}. In \S\ref{sec5} we draw some philosophical morals, in particular by focusing on the crucial differences between indeterminacy in QG and in non-relativistic QM. In \S\ref{sec6} we conclude. 

\section{Ontological Indeterminacy in Quantum Mechanics}\label{sec2}
The \textit{lack of value-definiteness} (LVD) indicates that quantum observables do not possess definite values at all times. In order to have a better grasp on LVD, following the existing literature (refs) we shall start by stating the general principle through which we can assign values to physical quantities based on the quantum state. In the standard formalism, this is done by the so-called Eigenstate-Eigenvalue Link (EEL),\footnote{Though note than some, most notably \cite{wallace2019orthodox}, disagree that EEL be part of the standard formalism. See \cite{gilton2016whence} for a defence of EEL as part of standard QM.} which can be stated as follows:

\begin{itemize}
    \item[\textbf{EEL:}] A physical system \textit{s} has a definite value \textit{v} of a given observable $\mathcal{O}$ \textit{if and only if} \textit{s} is in an eigenstate of $\mathcal{O}$.
\end{itemize}

\noindent Through the EEL, we can then give a classification of various cases where LVD seemingly emerges. The most accurate of such classification has arguably been given by \cite{calosi2019quantum}, who individuate three distinct features of the theory giving rise to LVD. These are (i) \textit{Incompatible Observables}, (ii) \textit{Superposition}, and (iii) \textit{Entanglement}. As regards to (i), consider two observables $\mathcal{O}_1$ and $\mathcal{O}_2$. These observables commute \textit{if and only if} $[\mathcal{O}_1,\mathcal{O}_2]=\mathcal{O}_1\mathcal{O}_2-\mathcal{O}_2\mathcal{O}_1$ = 0$ $. If they do not satisfy this constraint, they do not commute, and are called \textit{incompatible}. Since incompatible observables do not share the same eigenstates, if the system \textit{s} is in one such eigenstate of, say, $\mathcal{O}_1$, it follows that it does not have a definite value for $\mathcal{O}_2$ (and \textit{viceversa}). As regards to (ii), note that a linear combination $\ket{\psi}=q_1\ket{\phi_1}+q_2\ket{\phi_2}$ of different eigenstates $\ket{\phi_1}$ and $\ket{\phi_2}$ of an observable $\mathcal{O}$ is not always an eigenstate of $\mathcal{O}$. If a system $s$ is in $\ket{\psi}$, it follows that it does not have a definite value of $\mathcal{O}$. Finally, take (iii), \textit{entanglement}. Consider a system $s_{12}$ composed by $s_1$ and $s_2$ with corresponding Hilbert space $\mathcal{H}_{12}=\mathcal{H}_1\otimes\mathcal{H}_2$. $s_{12}$ may be in an eigenstate $\ket{\psi}$ of $\mathcal{O}_{12}$ that is neither an eigenstate of $\mathcal{O}_1$ nor an eigenstate of $\mathcal{O}_2$---with $\mathcal{O}_1$ and $\mathcal{O}_2$ defined on $\mathcal{H}_1$ and on $\mathcal{H}_2$ respectively. From this it follows that $s_1$ and $s_2$ will lack a definite value for both corresponding observables. Although there are crucial conceptual differences between these three cases (see \cite{calosi2019quantum} for an extensive discussion), for what matters to us here the result of applying EEL to each of them is the same, namely that one or more observables do not always possess a definite value.
%\footnote{priority between the cases?}

Defenders of \textit{quantum indeterminacy}, such as \cite{calosi2019quantum}, argue that LVD should be taken at face value as indicating that the world is \textit{ontologically indeterminate}.\footnote{See also: \cite{bokulich2014metaphysical,wolff2015spin,darby2010quantum,darby2014vague,torza2020quantum,David2021,corti2021yet,fletcher2021quantum}. For a critique, see \cite{glick2017against}, while, for an extensive review of the debate, see \cite{calosi2021quantumreview}.} Two distinct families of approaches have been proposed to account for this, which \cite{wilson2016there} calls \textit{meta-level} and \textit{object-level} views. According to the former, very roughly, indeterminacy is understood as worldly unsettledness between fully precise alternatives. On this view, ontological indeterminacy occurs whenever it is indeterminate which determinate state of affairs obtains. This view is meant to capture the phenomenon of indeterminacy \textit{modally}, and in a way not dissimilar from how we account for other notions such as possibility or necessity. According to the \textit{object-level} view, instead, indeterminacy is understood as the (determinate) obtainment of an indeterminate state of affairs. The crucial explanatory component of this view is then played by the definition of indeterminate state of affairs, which can potentially be given in various ways. The most discussed view, again developed by \cite{wilson2013determinable}, exploits the distinction between determinable and determinate properties. On this view, an indeterminate state of affairs is one where a given entity instantiate a determinable (e.g. \textit{red}) without instantiating a unique determinate (e.g. \textit{scarlet} or \textit{crimson}). The non-uniqueness requirement can in turn be satisfied in at least three ways: \textit{gappy} has it that no determinate is instantiated; \textit{glutty relativized} has it that more than one is instantiated, although each relative to something; \textit{glutty degree} has it that more than one is instantiated, each with a degree less than 1.\footnote{A major issue in the philosophical debate on \textit{ontological indeterminacy}, at least since \cite{evans1978can} famous paper, concerns the very coherence of this notion. We shall not enter this debate here, in large part because we believe that the models we just sketched offer ways to escape Evans' conclusion. By this, we do not mean to claim that the debate is over. Nonetheless, for what matters to us, we can simply assume that these accounts are consistent ways for understanding the notion of \textit{ontological indeterminacy}.}  
The argument leading from LVD to the existence of quantum indeterminacy is not necessarily tied to the EEL. For instance, ontological indeterminacy may also arise in other interpretations of QM which reject this link.\footnote{Most prominently, in the case of spontaneous collapse models, there have been several proposals on how to revise the EEL, all of which retain some indeterminacy. Examples include \cite{albert1996tails}'s \textit{fuzzy link}, \cite{lewis2016quantum}'s \textit{vague link}, and \cite{mariani2020non}'s \textit{degree link}.} However, it is common practice in the literature to start with the EEL so as to give the most clear explanation of the emergence of ontological indeterminacy. We shall come back to this issue later on, in \S5, when we will discuss the case of QG in more details.

\section{Loop Quantum Gravity}\label{sec3}
Our primary goal in this section is to explore the appearance of ontological indeterminacy in QG by dint of a simple model naturally arising in Loop Quantum Gravity \citep{Rovelli_2003,Livine_2004}. In particular, we will see that the indeterminacy emerging from these constructions is of the incompatible observables type, i.e.\ that two (or more) observables cannot both have well-defined values at the same time. This fact stems from a given system instantiating properties described by those operators that cannot be in an eigenstate at the same time. For our presentation, we draw in particular from \citep{Rovelli_2003,Livine_2004}.

Let us first remark on one of the fundamental features that we expect any theory of quantum gravity to realise: discreteness of spacetime. In particular, we expect that the spacetime described by a consistent quantum gravitational model will display a minimal length of some kind, and that this minimal length coincides with the Planck length. By minimal length here, we mean a length such that no observer can measure lengths shorter than that.\footnote{If the reader does not like talk of measurements in this context, they can remove measurement from the discussion and think about there being a minimal length. The same disclaimer applies to all other occurrences of operational language in this article.} Thus, if we take the Planck length to be our minimal length, no observer is allowed to measure lengths shorter than the Planck length. However, such a minimal length is apparently in contrast with Lorentz invariance, as we will see now. This contrast between minimal lengths and Lorentz invariance and its resolution will be the origin of ontological indeterminacy in these models. 

Let us start by briefly looking at why minimal length and Lorentz invariance are in apparent contrast. It is useful to start by remembering that one of the symmetry transformations of the Lorentz group are boosts,\footnote{Indeed, boosts are Lorentz transformations not involving spatial rotations.} which take an observer in a reference frame at rest to an observer in a reference frame moving with constant velocity with respect to the first. In a Lorentz invariant theory, such boosts will usually take the following form:
\beq
t' &=& \gamma (t - \frac{vx}{c^2}) \\
x' &=& \gamma (x - vt) \\
y' &=& y \\
z' &=& z
\eeq
where $\gamma = \frac{1}{\sqrt{1 - \frac{v^2}{c^2}}}$, $(t,x,y,z)$ and $(t',x',y',z')$ are sets of coordinates in two different reference frames moving with relative velocity $v$ in direction $x$, and $c$ is the speed of light.

The critical feature of these transformations for this article is that they give rise to two of the most celebrated predictions of special relativity (which extend to any Lorentz invariant theory): length contraction and time dilation. In particular, what matters here is length contraction. By length contraction, we mean that, if we take a rod of length $l$ in a given reference frame, upon moving to the reference frame of an observer moving at constant velocity $v$ with respect to us, we find that our rod in this new reference frame is no more of length $l$, but of length $l' = \frac{1}{\gamma} l$. Thus, our rod got shorter when moving to a boosted reference frame. 

With this said, one can immediately see why Lorentz boosts, and thus Lorentz invariance, are not compatible with a minimal length. Take an observer in a given reference frame who measures a given length. Let us also stipulate that the result of their measurement is the minimal length $l$, which we identify with the Planck length $l_{Planck}$ for simplicity.\footnote{We are bracketing potential issues regarding the practical feasibility of such measurements.} Let us now move to a boosted reference frame. In this new reference frame, by length contraction, a boosted observer performing our original observer's measurement will not get $l_{Planck}$ as a result, but $l' = \frac{1}{\gamma} l_{Planck}$, i.e. less than $l_{Planck}$. Thus, we have seen a way to measure a length shorter than the minimal length, which is, of course, incompatible with the definition itself of minimal length. Thus, Lorentz invariance and minimal lengths are incompatible.

The argument, as stated, is certainly less than conclusive and can be resisted at various points. Let us briefly mention only one: gravity is not (globally) Lorentz invariant, but only locally Lorentz invariant. While this fact is undoubtedly true, let us observe two points in response. (i) violations of local Lorentz invariance would be troubling nonetheless, and in any case, (ii) one can reformulate the above scenario by taking a quantum gravitational spacetime whose classical limit is flat spacetime and considering small scale quantum effects in this classical background, which would then display the above-mentioned features.\footnote{Indeed, this is the approach taken in \cite{Rovelli_2003} to analyse the apparent contradiction between Lorentz invariance and minimal length and show that it does not arise in Loop Quantum Gravity.} Moreover, besides the various issues with the argument presented above, what is instructive is to see how, concretely, we can avoid situations such as those described above in the context of a theory of quantum gravity such as Loop Quantum Gravity. Note that nothing in this article depends on the air-tightness of the argument presented above, but only on the features of Loop Quantum Gravity that avoid its conclusion, which are independent of the argument's assumptions.

To start, let us briefly recast the above scenario in a way more natural for Loop Quantum Gravity \citep{rovelli_2004,rovelli_vidotto_2014}. Loop Quantum Gravity is an approach to quantum gravity that starts by recasting General Relativity in terms of a new set of variables (Ashtekar's variables) which put General Relativity in a form close to SU(2) Yang-Mills theory. From this point of view, one then attempts to quantise gravity in the standard way that one would quantise any field theory. The result of this quantisation procedure is then Loop Quantum Gravity. While we will not be concerned with the details of Loop Quantum Gravity's formulation, it is important to note that in Loop Quantum Gravity, discreteness of spacetime quantities is most naturally seen in the discreteness of the spectra of the area and volume operators, rather than in an explicit minimal length. We will thus focus on area operators, though note that nothing substantial changes from the discussion in terms of a minimal length. The spectrum of the area operator $\mathcal{A}$ is the following:
\beq
\braket{\mathcal A} = 8\pi\gamma l^2_{Planck} \sqrt{j(j+1)},
\eeq
where $\braket{\mathcal A}$ is an area eigenvalue, $\gamma$ is a free parameter,\footnote{Not to be confused with the $\gamma$ in the Lorentz boosts.} $l_{Planck}$ is Planck's length and $j$ are irreducible representations of the group $SU(2)$, which means that $j=0,\frac{1}{2},1,\frac{3}{2},2,\dots$. It is immediate to see that the spectrum of the area operator $\mathcal{A}$ is discrete since the $j$s, being representations of $SU(2)$, are discrete. This fact also implies that there is a minimal area, corresponding to the smallest non-zero eigenvalue of $\mathcal{A}$.

To get our minimal length problem, one can repeat the steps described above and consider what a Lorentz boosted observer would observe if they were to measure the area. Naively, we would expect a contraction of the area and thus to have a conflict between Lorentz invariance and minimal area in Loop Quantum Gravity. This description, however, is not the whole story. Let us be slightly more precise about what is going on in this scenario. An observer is measuring the area operator $\mathcal{A}$, thus projecting into an eigenstate of $\mathcal{A}$.\footnote{We follow here usual practice in treating measurements as projections and bypass worries about the measurement problem, as they will not be relevant to our discussion here.} They then compare their result with that of a second observer, who measures a different operator, $\mathcal{A'}$, which is the result of applying a Lorentz boost to $\mathcal{A}$. Now, this comparison is straightforward only if $\mathcal{A}$ and $\mathcal{A'}$ commute, otherwise we cannot compare the two since they would not have well-defined values at the same time because they would not have eigenstates in common. Luckily, as shown in \cite{Rovelli_2003}, $\mathcal{A}$ and $\mathcal{A'}$ do not commute:
\beq
[\mathcal{A},\mathcal{A'}] \neq 0~.
\eeq
Thus, when one measures $\mathcal{A}$ and then tries to see if $\mathcal{A'}$ will show a contracted area smaller than the minimal one by length contraction, their efforts will ultimately crash against the fact that $\mathcal{A'}$ does not show a well-defined for area since the system is not in an eigenstate of $\mathcal{A'}$. Thus, it does not make sense to ask what is the area measured by the observer measuring $\mathcal{A'}$, as there is no definite answer to this question since we find a superposition of eigenstates of $\mathcal{A'}$\footnote{Which has the same spectrum of $\mathcal{A}$, but being on a superposition of eigenstates of  } whenever we observe an eigenstate of $\mathcal{A}$.  While $\mathcal{A}$ and $\mathcal{A'}$ have the same spectrum, eigenstates of $\mathcal{A}$ correspond to superpositions of eigenstates of $\mathcal{A'}$. In this way, within Loop Quantum Gravity, we avoid the apparent conflict between a minimal area and Lorentz invariance because, speaking somewhat operationally, in the boosted frame, one measures a statistical distribution of discrete eigenvalues of area, rather than a single contracted eigenvalue. Interestingly enough, as observed by \cite{Rovelli_2003}, this result relies essentially\footnote{Modulo various technical subtleties which are not relevant for the argument in this section.} on the same mechanism by which quantised angular momentum is compatible with invariance under rotations in Quantum Mechanics. In both cases, it is the non-commutativity of certain operators connected by the relevant symmetry which allows us to implement the symmetry and preserve their spectrum at the same time by making it impossible to have an eigenstate for the two operators at the same time. Without a single eigenstate for both operators, the symmetry-related operators (for example, by Lorentz boosts) will not both have meaningful information about areas at the same time, let alone about an area smaller than the minimal one. Since, however, the claim that Lorentz boosts and minimal area were in tension relied crucially on the comparison between what an operator at rest and one Lorentz boosted could tell about area, the statement of this tension is itself impossible.

It is immediate now to see the connection with ontological indeterminacy. Since incompatible observables are formally represented as non-commuting operators, and since incompatible observables are taken to be one of the main sources of ontological indeterminacy in QM, it seems that also, in this case coming from LQG, we should have ontological indeterminacy. However, before moving to consider the application of ontological indeterminacy to quantum gravity more in detail, let us consider a second example of the appearance of ontological indeterminacy in quantum gravity, this time coming from string theory. We will do this in the next section. 

\section{String Theory}\label{sec4}
In this section, we are going to look at how ontological indeterminacy appears within string theory. Before starting, let us make two technical disclaimers for the ensuing discussion. As with any quantum field theory, string theory has two regimes: perturbative and non-perturbative. After the development, on one side, of holography and AdS/CFT \citep{maldacena1999large}, and, on the other, of M-theory \citep{witten1995string} and F-theory \citep{vafa1996evidence}, much progress has been made in the non-perturbative regime of String Theory, which, however, remains far from being satisfactorily under control. In this section, then, we will focus on the perturbative sector of String Theory, the so-called \textit{perturbative String Theory}, which is, by now, well-known.\footnote{For a book-lenghth treatement, see \cite{green2012superstring}, on which we draw for our discussion.} Moreover, for ease of exposition, we will concentrate just on Bosonic String Theory. The extension of our discussion to Superstring Theory is straightforward.\footnote{One needs to add fermions on the worldsheet by enforcing supersymmetry.} 

The fundamental objects of perturbative string theory are one dimensional objects, the (closed and open) strings. As the dynamics of a point-particle can be represented via its one-dimensional \textit{world-line}, the dynamics of a string can be represented via its two-dimensional \textit{world-sheet}, which we call $ \Sigma$. The action $\mathcal S$ of perturbative string theory, called the Polyakov action, is a generalisation of the action of a point-particle, and can be written as follows:
\beq\label{eq:poly}
\mathcal S_{Poly} \left[h,X\right]= \frac{1}{L_s^2} \int d^2 \sigma \sqrt{-h}~h^{\alpha\beta} \partial_\alpha X^\mu \partial_\beta X^\nu g_{\mu\nu} \left(X^\rho \right)~,
\eeq
where $L_s$ is the string length, $\sigma^0 = \tau$ and $\sigma^1= \sigma$ are the world-sheet coordinates, $h^{ \alpha \beta}$ and $h$ are respectively the inverse metric and the determinant of the world-sheet metric $h_{ab}$, which describes the geometry of the world-sheet. $X^ \mu \left( \sigma \right)$ is a map between the string world-sheet and the target space, i.e.\ the spacetime in which the string propagates, while $g_{ \mu \nu} \left(X \right)$ is at the same time the coupling constant of the string interactions and the metric of target space. 

The Polyakov action \eqref{eq:poly} has, \textit{prima facie}, two degrees of freedom, namely the worldsheet metric $h_{\alpha\beta}$ and the embedding coordinates $X^\mu$. Moreover, the action \eqref{eq:poly} has two local symmetries (diffeomorphisms and Weyl symmetry), and one global symmetry, inherited by target space (Poincar\'e invariance). In order to make sense of the quantum mechanical description of \eqref{eq:poly}, i.e.\ in order to define a path integral, write down correlation functions and compute scattering amplitudes, we need to fix a gauge for the local symmetries. A complete review of the construction of the path integral for the Polyakov action \eqref{eq:poly} goes beyond the scope of this paper. We limit ourselves to the result of the quantisation procedure. In order to write down a path integral for theories with local symmetries, one first needs to gauge fix. Since the worldsheet metric can be completely gauge fixed it does not have any propagating degrees of freedom and thus it is not part of the physical content of the theory.\footnote{This is why we do not consider the worldsheet metric below.} The upshot of the quantisation procedure is that one trades the path integral over the metric $h_{\alpha\beta}$ with a path integral over the new (ghost) fields $b,~\bar b, ~c$. One can write correlation functions of $N$ (gauge invariant) operators in the following way:
\beq\label{eq:corr}
\braket{\braket{V_1\dots V_N}} &=& \int \frac{\mathcal D X^\mu ~\mathcal D h_{\alpha\beta}}{\mbox{Vol}\left(\mbox{gauge}\right)}\left(V_1\dots V_N \right)e^{-S_{Poly} \left[h,X\right]} \\ 
&=& \int \mathcal D X^\mu ~\mathcal D b ~\mathcal D \bar b ~\mathcal D c\left(V_1\dots V_N \right) e^{-S_{matter} \left[X\right]- S_{ghost}\left[b,\bar b, c\right]}
\eeq
where $S_{matter} \left[X\right]$ is the \textit{matter action} and $S_{ghost}\left[b,\bar b, c\right]$ is the \textit{ghost action}. In the following we would like to review the quantisation of the space of solutions of the gauge fixed Polyakov action, i.e.\ the possible perturbative vacua of quantum theory around which we build perturbation theory and evalutate the correlation functions \eqref{eq:corr} via the path integral. We do this to highlight where ontological indeterminacy arises in String Theory. Let us focus on the matter action $S_{matter} \left[X\right]$. By a proper choice of gauge, the so-called conformal gauge,\footnote{The conformal gauge is the gauge in which the worldsheet metric $h_{\alpha \beta}$ is flat, i.e.\ $h_{\alpha \beta}=\delta_{\alpha\beta}$. The following is also written in a new set of worldsheet coordinates: $\omega = t+i\sigma$ and $\bar \omega = t-i\sigma$ and $(\tau, \sigma) \rightarrow (\omega,\bar\omega)$.} one can write the equation of motion derived from $\mathcal S_{\mbox{matter}}$ in the following way:
\beq\label{eq:eomX}
\partial \bar\partial X^\mu = 0
\eeq
where $\partial$ and $\bar \partial$ are derivatives with respect to $\omega$ and $\bar \omega$ respectively. A general solution of \eqref{eq:eomX} can be written in the form $X^\mu(\omega,\bar\omega) = X_L^\mu (\omega) +X_R^\mu (\bar \omega)$, where $X_L^\mu (\omega)$ and $X_R^\mu (\bar \omega)$ are respectively left and right moving modes of the string, and are holomorphic functions of $\omega$ and $\bar \omega$. Since there are two types of fundamental objects in perturbative String Theory, open and closed strings, we need to solve this equation of motion for both of them. 
\begin{itemize}
\item[] \textbf{Closed Strings}: they are defined by asking periodicity on $\omega$ and $\bar \omega$, i.e.\ $X^\mu \left(\omega + 2\pi i, \bar \omega-2\pi i\right) = X^\mu \left(\omega, \bar \omega
\right)$. 
\begin{figure}[h!]
\centering
\includegraphics[scale=1]{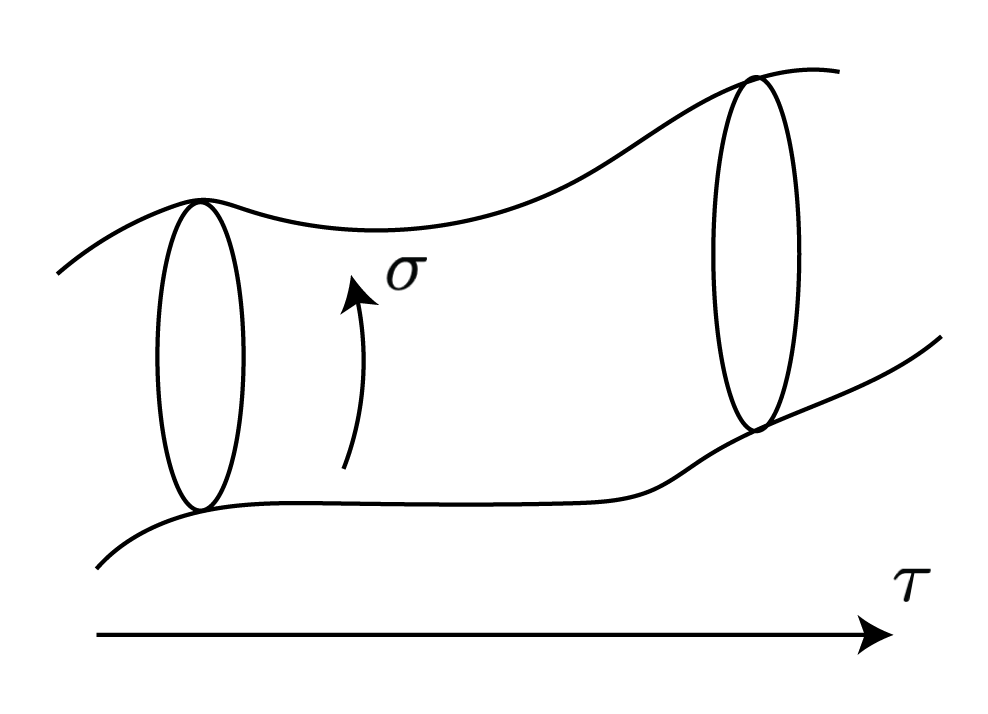} 
\caption{A schematic picture of a closed string propagating in spacetime. $\sigma$ represents the periodic spatial coordinate of the worldsheet, while $\tau$ represents its time coordinate.}
\label{fig:chiusa}
\end{figure} 
A general solution for left and right moving modes of the string can be written in the following form
\beq\label{eq:Xclosed}
X_L ^\mu (\omega) &=& \frac{1}{2}X_0^\mu -\frac{i}{2}\alpha' P^\mu \omega + i \sqrt{\frac{\alpha'}{2}}\sum_{n\neq 0} \frac{\alpha_n}{n}e^{n\omega}\\
X_R^\mu (\omega) &=& \frac{1}{2}X_0^\mu -\frac{i}{2}\alpha' P^\mu \bar\omega + i \sqrt{\frac{\alpha'}{2}}\sum_{n\neq 0} \frac{\bar\alpha_n}{n}e^{n\bar\omega}
\eeq
which are chosen in order to solve the equation of motion \eqref{eq:eomX} and obey the periodicity condition. As common in analytic mechanics, one can also write down the momentum density, i.e.\ the conjugate variable of $X^\mu$:\footnote{The momentum density $p$ in analytic mechanics is defined as the functional derivative of the lagrangian with respect to the temporal derivative of $q$, i.e.\ $p := \frac{\partial \mathcal L}{\partial \dot q}$, which generalises to our setup as $\mathcal P_\mu := \frac{\partial \mathcal L}{\partial \partial_t X^\mu}$.} 
\beq\label{eq:Pclosed}
\mathcal P_\mu = \frac{i}{2\pi \alpha'} \left(\partial X_\mu+\bar \partial X_\mu\right) = \dots
\eeq
Having $X^\mu$ and $\mathcal P_\mu$, one can apply the rules of canonical quantisation, i.e.\ demand that  $\left[X^\mu(t,\sigma), \mathcal P_\nu (\tau,\sigma')\right] = i~ \eta^\mu{}_\nu ~\delta \left(\sigma-\sigma'\right)$. Using the explicit expansions of $X^\mu$ and $\mathcal P_\mu$ given by equations \eqref{eq:Xclosed} and \eqref{eq:Pclosed}, one gets the following commutation relations for the expansion coefficients of the $X^\mu$s: 
\beq
\label{eq:comm1} 	\left[X_0^\mu, P^\nu\right] &=& i~ \eta^{\mu\nu} \\
\label{eq:comm2} 	\left[\alpha_n^\mu , \alpha_m ^\nu \right] &=& n~ \eta^{\mu\nu} \delta_{n+m}\\
\label{eq:comm3} 	\left[\bar \alpha_n^\mu , \bar \alpha_m ^\nu \right] &=& n~ \eta^{\mu\nu} \delta_{n+m}\\
\label{eq:comm4} 	\left[\alpha_n^\mu , \bar \alpha_m ^\nu \right] &=& 0
\eeq
The Hilbert space of the theory for closed strings is $\mbox{Span} \left\{\alpha_{-n_1}^{\mu_1} \dots \alpha_{-n_k}^{\mu_k}\bar\alpha_{-m_1}^{\nu_1} \dots\bar \alpha_{-m_q}^{\nu_q}\right\}\ket{0,p}$.\footnote{The choice of the vacuum state $\ket{0,P}$ and the properties of this Hilbert space go beyond the scope of this paper.}
\item[]\textbf{Open Strings}: the pecularity of the open string sector is that, together with the equation of motion \eqref{eq:eomX}, the open string should obey also a boundary constraint equation, namely $\delta X_\mu \partial_\sigma X^\mu = 0 |_{\sigma = 0,\pi}$. 
\begin{figure}[h!]
\centering
\includegraphics[scale=1]{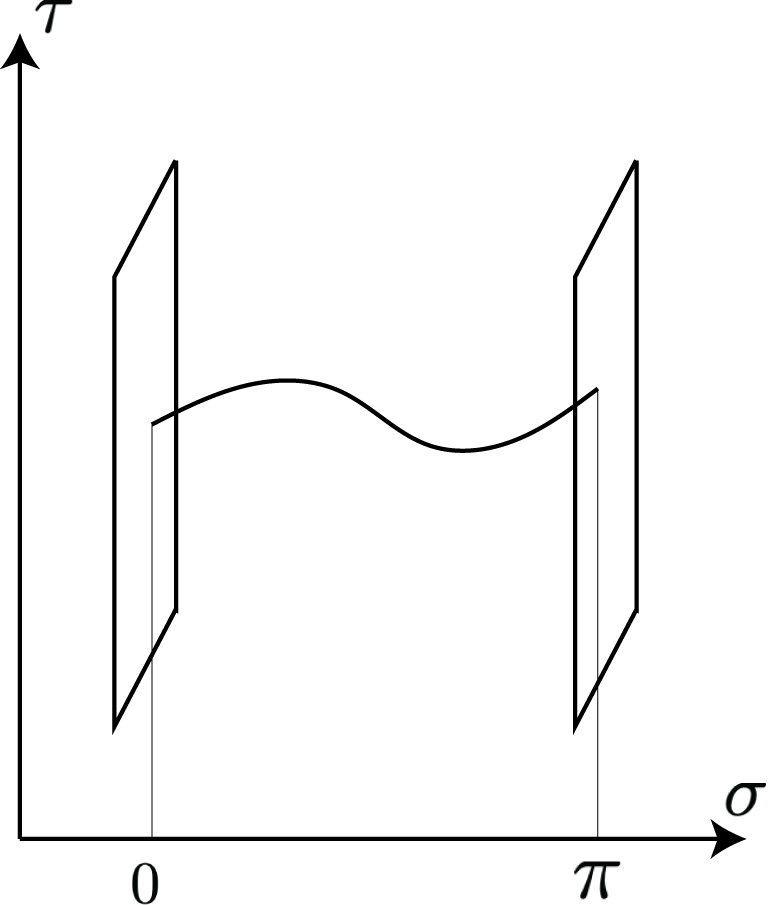} 
\caption{A schematic picture of an open string propagating in spacetime with its endpoints constrained on two D-Branes which give its boundary conditions. $\sigma$ represents the spatial coordinate of the worldsheet, while $\tau$ represents its time coordinate.}
\label{fig:aperta}
\end{figure}
The solution of the equation of motion in the open string sector heavily depends on the choice of the boundary conditions, i.e.\ on the solution of the boundary constraint equation. The most well-known choices of boundary conditions are Neumann's ($\partial_\sigma X^\mu |_{\sigma = 0,\pi}$) and Dirichlect's ($\delta X^\mu |_{\sigma = 0,\pi}$). Physically, Neumann boundary conditions (N) entail that there is no flow of momentum off the boundary, while Direchlet boundary conditions (D) entail that the endpoint of the strings cannot move.\footnote{This discussion of boundary conditions for the open string sector naturally leads to the definition of D-branes, which, however, we do not discuss in this paper.} As in the case of closed strings, also for the open string sector one can write the solution to the equation of motion in the form $X^\mu(\omega,\bar\omega) = X_L^\mu (\omega) +X_R^\mu (\bar \omega)$, and a general solution is:
\beq
X_L ^\mu (\omega) &=& \frac{1}{2}X_L^\mu -i\alpha' P_L^\mu \omega + i \sqrt{\frac{\alpha'}{2}}\sum_{n\neq 0} \frac{\alpha_n}{n}e^{n\omega}\\
X_R^\mu (\omega) &=& \frac{1}{2}X_R^\mu -i\alpha' P_R^\mu \bar\omega + i \sqrt{\frac{\alpha'}{2}}\sum_{n\neq 0} \frac{\bar\alpha_n}{n}e^{n\bar\omega}
\eeq
Now, one needs to apply boundary conditions which give us constraints on $X_L ^\mu (\omega)$ and $X_R ^\mu (\omega)$. There are four possible choices of boundary conditions: NN, ND, DN, DD.\footnote{Recall that we have to fix the boundary conditions at both ends of the string, i.e.\ at $\sigma = 0,\pi$, which means two choices of boundary conditions (not necessarily the same on both ends of the string). For example, when we write NN, we mean Neumann boundary conditions at both ends of the string. We can recast these four possibilities in matrix form via the Chan-Paton matrix.} Without delving into unnecessary details, the upshot of imposing boundary conditions is that, for open strings, one only has one set of oscillators, i.e.\ $\alpha=\bar\alpha$. The commutation properties of the oscillator modes heavily depend on the choice of boundary conditions. For instance, we can have half-integer oscillator modes for mixed boundary conditions, i.e.\ ND and DN. By imposing the canonical quantisation procedure also to the open string sector, one obtains commutation relations among oscillator modes similar to those obtained for the closed string sector, i.e.\ \eqref{eq:comm1}-\eqref{eq:comm4}. 
\end{itemize}
A similar analysis can also be done for the ghost sector of the theory by studying the quantisation of the space of solutions of $\mathcal S_{ghost} [b,~\bar b,~c]$, which, however, is not relevant for our argument here. We are instead interested in a physical interpretation of the commutation relations \eqref{eq:comm1}-\eqref{eq:comm4}. For instance, in the closed string expansion \eqref{eq:Xclosed}\footnote{A similar interpretation of the modes of the expansion also exists for open strings.} \dots $X_0^\mu$ represents the centre of mass position of the string, $P_\mu$ represents its centre of mass momentum, while $\alpha$ and $\bar \alpha$ represents its higher harmonic modes. The commutation relations \eqref{eq:comm1}-\eqref{eq:comm4}, thus, entail that there are non-trivial commutation relations between these observables. 

Again, as in section \S\ref{sec3}, we see that the appearance of the non-commutativity of certain observables leads to non-trivial commutation relations, which can be naturally understood, as we have learned in QM, as meaning that certain observables are incompatible. In the next section, we will look more in detail at some of the metaphysical implications of this connection. 

\section{Interpreting Non-Commutativity in QG}\label{sec5}
The two cases we presented rely on the appearance of non-commutativity to show that certain physical quantities cannot jointly be assigned a definite value in QG. These quantities, to recall, are represented by the Lorentz boosted and unboosted area operators to account for the relation between minimal area and Lorentz invariance (in LQG), and by the position, momentum, and higher harmonic modes operators (in ST). Given the structural analogies between these cases and the standard case of ontological indeterminacy in QM (discussed in \S\ref{sec2}), we are now in the position to evaluate the status of accounts of ontological indeterminacy in QG as well. This task goes well beyond the scope of this paper, and yet we shall indicate some important analogies and disanalogies that may serve programmatically for future developments. 

First, consider that the standard quantum mechanical argument heavily relied on the eigenstate-eigenvalue link (EEL). So, for a similar line of reasoning to be applied here, we should first of all establish whether something like the EEL can be stated in QG. To the best of our knowledge, there is no explicit discussion of the EEL in QG. Nonetheless, there are two main considerations that seem to suggest that the EEL is both (i) preserved, and (ii) preserved in its standard form. As for (i), notice that the main role of the EEL is to map the abstract formalism onto physical states, so as to understand how these can be empirically detected.\footnote{Bracketing here concerns such as those of \cite{wallace2019orthodox}.} Therefore, it is natural to expect that empirical adequacy would require to assume something like the EEL. As for (ii), approaches to QG are usually meant to be continuous with standard QM, at least when it comes to basic features of the quantum formalism such as the EEL, and hence there is no need to believe (unless we have independent reasons) that a modified link would be needed. For these reasons, we take it that the acceptance of the EEL is a safe assumption.

Let us now comment on the relevance of these constructions for ontological indeterminacy. The crucial point for this article is that both in LQG and ST, the consistency of the quantum theory of gravity relies crucially on the claim that certain operators cannot all represent determinate properties at the same time. This is the same situation as that of ontological indeterminacy from incompatible observables described in \S\ref{sec2}. We have two properties and a physical theory that forces one of them to be metaphysically indeterminate whenever the other is determinate, on pains of inconsistency with the theory itself. The incompatibility of these properties, as usual in any quantum theory, is encoded in the non-zero commutator between the operators representing these properties. Indeed, in LQG, we even have an explicit analogy with angular momentum, which makes particularly evident the connection with analyses of ontological indeterminacy developed in quantum mechanics. Also in string theory, the fact that we are basically dealing with non-commutativity of position and momentum operators is immediately analogous with the standard quantum mechanical case. In QM, this kind of non-commutativity is precisely the type of phenomenon falling under the rubric of incompatible observables. All that we need is the extension of this claim to quantum gravity, and the structural similarity between the two cases suggests that this extension should be accepted. Indeed, in all these cases, all one is relying on is the non-commutativity of certain observables. Moreover, in all these cases, we interpret this non-commutativity in the same way, as the impossibility of instantiating two properties determinately at the same time. Then a refusal of treating the quantum gravity case as instantiating incompatible observables indeterminacy appears at best misguided, at worst inconsistent with the attitude taken towards the quantum mechanical case, be it that of angular momentum or that of position and momentum. Nevertheless, at the end of the day, these considerations stand and fall with the acceptance of the argument from non-commutativity of observables to ontological indeterminacy. However, our goal here was not to show that ontological indeterminacy is inevitable in quantum gravity but only that it is as natural as in quantum mechanics. Indeed, the two cases share some crucial features that allow us to extend the quantum mechanical analysis of indeterminacy to quantum gravity. It is this goal we claim to have reached.
%specificare che indeterminacy richiede anche EEL; espandere sul fatto che se si prende una certa attitudine su EEL e MI in QM allora bisogna prenderla anche in QG perchè i due casi sono analoghi; spezzare in due paragrafi, uno col grosso di questo argomento che c'è già su non commutativity, e uno che spieghi bene la connessione con QM

Let us now look at an aspect where QM and QG, at least in the examples we have considered here, diverge in the way they instantiate metaphysical indeterminacy. As we discussed in this section, and reviewed more extensively in\S\ref{sec2}, non-commutativity, along with the EEL, entails that certain physical quantities are indefinite. In the case of QM, the relevant quantities can be given, at least to some extent, a clear physical meaning. The same is not as straightforward for QG, for as we discussed they are meant to represent the geometrical structure of spacetime. While it could be accepted that a certain object possesses an indeterminate location, it seems much harder to believe that the \textit{regions} themselves are indeterminate. And yet, despite its counterintuitive nature, this claim is supported (though in quite distinct ways) by both approaches to QM discussed in this paper. This seems to suggest that the possibility of spacetime being fundamentally indeterminate has to be taken seriously. 
%connessione/compare and contrast con location in QM? Vedere Claudio

\section{Conclusions}\label{sec6}
The goal of this paper was to argue that there is a deep conceptual continuity between QM and QG in considerations regarding the so-called \textit{lack of value definiteness} for physical quantities. We have shown that on the two most developed approaches to QG, namely Loop Quantum Gravity and String Theory, the observables representing the geometric structure of spacetime are non-commuting ones. By building upon the standard reasoning leading to objective indeterminacy in standard QM, we have argued that these quantities can also be considered indeterminate. These results points towards the possibility that spacetime is fundamentally indeterminate according to QG.

\bibliography{Bibliography}

\begin{thebibliography}{}

\bibitem[\protect\citeauthoryear{Albert and Loewer}{Albert and
  Loewer}{1996}]{albert1996tails}
Albert, D.~Z. and B.~Loewer (1996).
\newblock Tails of schr{\"o}dinger’s cat.
\newblock In {\em Perspectives on quantum reality}, pp.\  81--92. Springer.

\bibitem[\protect\citeauthoryear{Bokulich}{Bokulich}{2014}]{bokulich2014metaphysical}
Bokulich, A. (2014).
\newblock Metaphysical indeterminacy, properties, and quantum theory.
\newblock {\em Res Philosophica\/}~{\em 91\/}(3), 449--475.

\bibitem[\protect\citeauthoryear{Calosi and Mariani}{Calosi and
  Mariani}{2020}]{calosi2020quantum}
Calosi, C. and C.~Mariani (2020).
\newblock Quantum relational indeterminacy.
\newblock {\em Studies in History and Philosophy of Science Part B: Studies in
  History and Philosophy of Modern Physics\/}~{\em 71}, 158--169.

\bibitem[\protect\citeauthoryear{Calosi and Mariani}{Calosi and
  Mariani}{2021}]{calosi2021quantumreview}
Calosi, C. and C.~Mariani (2021).
\newblock Quantum indeterminacy.
\newblock {\em Philosophy Compass\/}~{\em 16\/}(4), e12731.

\bibitem[\protect\citeauthoryear{Calosi and Wilson}{Calosi and
  Wilson}{2019}]{calosi2019quantum}
Calosi, C. and J.~Wilson (2019).
\newblock Quantum metaphysical indeterminacy.
\newblock {\em Philosophical Studies\/}~{\em 176\/}(10), 2599--2627.

\bibitem[\protect\citeauthoryear{Corti}{Corti}{2021}]{corti2021yet}
Corti, A. (2021).
\newblock Yet again, quantum indeterminacy is not worldly indecision.
\newblock {\em Synthese\/}, 1--21.

\bibitem[\protect\citeauthoryear{Darby}{Darby}{2010}]{darby2010quantum}
Darby, G. (2010).
\newblock Quantum mechanics and metaphysical indeterminacy.
\newblock {\em Australasian Journal of Philosophy\/}~{\em 88\/}(2), 227--245.

\bibitem[\protect\citeauthoryear{Darby}{Darby}{2014}]{darby2014vague}
Darby, G. (2014).
\newblock Vague objects in quantum mechanics?
\newblock In {\em Vague objects and vague identity}, pp.\  69--108. Springer.

\bibitem[\protect\citeauthoryear{Evans}{Evans}{1978}]{evans1978can}
Evans, G. (1978).
\newblock Can there be vague objects.
\newblock {\em Analysis\/}~{\em 38\/}(4), 208.

\bibitem[\protect\citeauthoryear{Fletcher and Taylor}{Fletcher and
  Taylor}{2021}]{fletcher2021quantum}
Fletcher, S.~C. and D.~E. Taylor (2021).
\newblock Quantum indeterminacy and the eigenstate-eigenvalue link.
\newblock {\em Synthese\/}, 1--32.

\bibitem[\protect\citeauthoryear{Gilton}{Gilton}{2016}]{gilton2016whence}
Gilton, M.~J. (2016).
\newblock Whence the eigenstate--eigenvalue link?
\newblock {\em Studies in History and Philosophy of Science Part B: Studies in
  History and Philosophy of Modern Physics\/}~{\em 55}, 92--100.

\bibitem[\protect\citeauthoryear{Glick}{Glick}{2017}]{glick2017against}
Glick, D. (2017).
\newblock Against quantum indeterminacy.
\newblock {\em Thought: A Journal of Philosophy\/}~{\em 6\/}(3), 204--213.

\bibitem[\protect\citeauthoryear{Green, Schwarz, and Witten}{Green
  et~al.}{1987}]{green2012superstring}
Green, M.~B., J.~H. Schwarz, and E.~Witten (1987).
\newblock {\em Superstring theory}.
\newblock Cambridge university press.

\bibitem[\protect\citeauthoryear{Lewis}{Lewis}{2016}]{lewis2016quantum}
Lewis, P.~J. (2016).
\newblock {\em Quantum ontology: A guide to the metaphysics of quantum
  mechanics}.
\newblock Oxford University Press.

\bibitem[\protect\citeauthoryear{Livine and Oriti}{Livine and
  Oriti}{2004}]{Livine_2004}
Livine, E.~R. and D.~Oriti (2004, Jun).
\newblock About lorentz invariance in a discrete quantum setting.
\newblock {\em Journal of High Energy Physics\/}~{\em 2004\/}(06), 050–050.

\bibitem[\protect\citeauthoryear{Maldacena}{Maldacena}{1999}]{maldacena1999large}
Maldacena, J. (1999).
\newblock The large-n limit of superconformal field theories and supergravity.
\newblock {\em International journal of theoretical physics\/}~{\em 38\/}(4),
  1113--1133.

\bibitem[\protect\citeauthoryear{Mariani}{Mariani}{2020}]{mariani2020non}
Mariani, C. (2020).
\newblock Non-accessible mass and the ontology of grw.
\newblock {\em arXiv preprint arXiv:2010.13706\/}.

\bibitem[\protect\citeauthoryear{Rovelli}{Rovelli}{2004}]{rovelli_2004}
Rovelli, C. (2004).
\newblock {\em Quantum Gravity}.
\newblock Cambridge Monographs on Mathematical Physics. Cambridge University
  Press.

\bibitem[\protect\citeauthoryear{Rovelli and Speziale}{Rovelli and
  Speziale}{2003}]{Rovelli_2003}
Rovelli, C. and S.~Speziale (2003, Mar).
\newblock Reconcile planck-scale discreteness and the lorentz-fitzgerald
  contraction.
\newblock {\em Physical Review D\/}~{\em 67\/}(6).

\bibitem[\protect\citeauthoryear{Rovelli and Vidotto}{Rovelli and
  Vidotto}{2014}]{rovelli_vidotto_2014}
Rovelli, C. and F.~Vidotto (2014).
\newblock {\em Covariant Loop Quantum Gravity: An Elementary Introduction to
  Quantum Gravity and Spinfoam Theory}.
\newblock Cambridge University Press.

\bibitem[\protect\citeauthoryear{Schr{\"o}dinger}{Schr{\"o}dinger}{1935}]{schrodinger1935gegenwartige}
Schr{\"o}dinger, E. (1935).
\newblock Die gegenw{\"a}rtige situation in der quantenmechanik.
\newblock {\em Naturwissenschaften\/}~{\em 23\/}(49), 823--828.

\bibitem[\protect\citeauthoryear{Schroeren}{Schroeren}{2021}]{David2021}
Schroeren, D. (2021).
\newblock Quantum metaphysical indeterminacy and the ontological foundations of
  orthodoxy.
\newblock {\em Studies in History and Philosophy of Science\/}.

\bibitem[\protect\citeauthoryear{Torza}{Torza}{2020}]{torza2020quantum}
Torza, A. (2020).
\newblock Quantum metaphysical indeterminacy and worldly incompleteness.
\newblock {\em Synthese\/}~{\em 197\/}(10), 4251--4264.

\bibitem[\protect\citeauthoryear{Vafa}{Vafa}{1996}]{vafa1996evidence}
Vafa, C. (1996).
\newblock Evidence for f-theory.
\newblock {\em Nuclear Physics B\/}~{\em 469\/}(3), 403--415.

\bibitem[\protect\citeauthoryear{Wallace}{Wallace}{2019}]{wallace2019orthodox}
Wallace, D. (2019).
\newblock What is orthodox quantum mechanics?
\newblock In {\em Philosophers Look at Quantum Mechanics}, pp.\  285--312.
  Springer.

\bibitem[\protect\citeauthoryear{Wilson}{Wilson}{2016}]{wilson2016there}
Wilson, J. (2016).
\newblock Are there indeterminate states of affairs? yes.
\newblock In {\em Current controversies in metaphysics}, pp.\  105--119.
  Routledge.

\bibitem[\protect\citeauthoryear{Wilson}{Wilson}{2013}]{wilson2013determinable}
Wilson, J.~M. (2013).
\newblock A determinable-based account of metaphysical indeterminacy.
\newblock {\em Inquiry\/}~{\em 56\/}(4), 359--385.

\bibitem[\protect\citeauthoryear{Witten}{Witten}{1995}]{witten1995string}
Witten, E. (1995).
\newblock String theory dynamics in various dimensions.
\newblock {\em Nuclear Physics B\/}~{\em 443\/}(1-2), 85--126.

\bibitem[\protect\citeauthoryear{Wolff}{Wolff}{2015}]{wolff2015spin}
Wolff, J. (2015).
\newblock Spin as a determinable.
\newblock {\em Topoi\/}~{\em 34\/}(2), 379--386.

\end{thebibliography}

\end{document}